\documentclass[aps,prb,showpacs,floatfix,twocolumn,superscriptaddress]{revtex4-2}
\usepackage{graphicx}
\usepackage{dcolumn}
\usepackage{cancel}
\usepackage{bm}
\usepackage{tikz}
\usetikzlibrary{calc}
\usepackage{siunitx}
\usepackage{tablefootnote}
\usepackage{physics}
\definecolor{darkblue}{rgb}{0,0,.65}
\definecolor{darkgreen}{rgb}{0,0.5,0}
\usepackage{nicefrac}
\usepackage{booktabs}
\usepackage[%
  pdfauthor={Benedikt Placke},%
  pdfstartview=FitH,%
  breaklinks=true,%
  bookmarks=true,%
  colorlinks=true,%
  anchorcolor=black,%
  citecolor=red,
  filecolor=black,%
  menucolor=black,%
  urlcolor=blue,%
  linkcolor=red,%
 ]{hyperref}

\usepackage[english]{babel}
\usepackage{amssymb}
\usepackage{color, soul}
\usepackage{amsmath}
\usepackage{amstext}
\usepackage{latexsym}
\usepackage{mathtools}
\usepackage[ruled]{algorithm2e}
\usepackage{enumitem}

\begin{document}
\title{Large-deviation tails of critical order-parameter distributions}
\author{Jinhong Zhu}
\thanks{These two authors contributed equally to this paper.}
\affiliation{
Hefei National Laboratory for Physical Sciences at the Microscale and Department of Modern Physics, University of Science and Technology of China, Hefei 230026, China}

\author{Yihao Xu}
\thanks{These two authors contributed equally to this paper.}
\affiliation{
Hefei National Laboratory for Physical Sciences at the Microscale and Department of Modern Physics, University of Science and Technology of China, Hefei 230026, China}
\affiliation{Hefei National Laboratory, University of Science and Technology of China, Hefei 230088, China}
\author{Abbas Ali Saberi}
\affiliation{
School of Science, Constructor University, Campus Ring 1, 28759 Bremen, Germany}

\author{Youjin Deng}
\thanks{yjdeng@ustc.edu.cn}
\affiliation{
Hefei National Laboratory for Physical Sciences at the Microscale and Department of Modern Physics, University of Science and Technology of China, Hefei 230026, China}
\affiliation{Hefei National Laboratory, University of Science and Technology of China, Hefei 230088, China}

\begin{abstract}
Large-deviation tails of critical probability distributions provide a sensitive probe of universality beyond standard finite-size scaling. We study these tails for critical percolation and Fortuin--Kasteleyn Ising models on two-dimensional lattices, three-dimensional lattices, and complete graphs. We consider two rescaled order parameters: the magnetization-like variable $x_m=|M|/\langle |M|\rangle$, including a signed cluster-mass analogue for percolation, and the largest-cluster variable $x_C=C_1/\langle C_1\rangle$. For $x_m$, we test the expected stretched-exponential large-deviation tail and show that the same form applies to the percolation analogue. For $x_C$, guided by the exact complete-graph result and scaling arguments, we propose universal scaling forms for both tails of the cumulative distribution and test them by extensive Monte Carlo simulations. In the complete-graph FK-Ising model, the left tail is governed by rare configurations with percolation-like scaling rather than by the typical Ising scaling. Our results show that the tails of order-parameter distributions reveal universal features of critical fluctuations that are not captured by averaged observables alone.
\end{abstract}

\maketitle 
           
\section{Introduction}
\label{sec:intro}

At a continuous phase transition, fluctuations are not merely large; they are organized across all length scales. A natural question is therefore not only which critical exponents describe typical fluctuations, but also how universality is encoded in the rare tails of critical observables. These tails measure the cost of producing atypically large or atypically small macroscopic fluctuations, and are naturally described in the language of large-deviation theory~\cite{Touchette2009}. From this viewpoint, the far tails of a critical probability distribution reveal how the critical fixed point penalizes coherent order-parameter fluctuations beyond the typical scaling regime. Identifying the universal and model-dependent parts of this tail structure has recently become an active problem in the theory of critical phenomena~\cite{Balog2022FRG,stella25,Balog2025SciPost,Camia2025PowerLaw}.

The conventional description of universality starts from finite-size scaling. For a finite system of linear size $L$, the singular part of an observable $\mathcal{O}$ obeys $\mathcal{O}(t,L)=L^{y_{\mathcal O}}\widetilde{O}(tL^{y_t})$, where $t$ is the reduced temperature, $y_t$ is the thermal exponent, $y_{\mathcal O}$ is the scaling exponent of the observable, and $\widetilde{O}$ is a scaling function. For the total magnetization $M$, the relevant exponent is the magnetic renormalization-group exponent $y_h$, so that at criticality $\langle M^k\rangle\propto L^{k y_h}$. A standard dimensionless quantity is the Binder ratio~\cite{Binder1981},
\begin{equation}
Q=
\frac{\langle M^2\rangle^2}{\langle M^4\rangle},
\label{eq:binder}
\end{equation}
which takes a universal value at the critical point and is widely used for locating criticality and extracting exponents. In this work, we focus instead on the full probability distribution of the order parameter, especially its asymptotic tails.

At criticality, the magnetization distribution satisfies $L^{y_h}P(M,L)=\widetilde{P}(M/L^{y_h})$. Equivalently, writing $N=L^d$ and $y_h^*=y_h/d$, one may use the scaling variable $x=M/N^{y_h^*}$. The large-$|x|$ behavior of $\widetilde{P}(x)$ then probes the large-deviation structure of critical magnetization fluctuations.

The mean-field limit gives the simplest example. In the infinite-dimensional limit, the critical Ising model is governed by the Landau--Ginzburg action $S[\phi]=N(r\phi^2+c\phi^4)$. At criticality, $r=0$, and the rescaled variable $x=N^{1/4}\phi$ has the quartic limiting distribution
\begin{equation}
P(x)\propto \exp(-c x^4).
\label{eq:phi4pdf}
\end{equation}
In this mean-field case the stretched-exponential form describes the limiting distribution itself, rather than only its far tail. The corresponding mean-field description of percolation is the $\phi^3$ theory, giving a leading tail of the form $\exp(-c x^3)$.

For percolation, corresponding to the $Q\to1$ Potts limit, there is no microscopic spin variable. To construct a magnetization-like observable, we use a signed cluster-mass field. For a given percolation configuration, let $C_a$ denote the size of cluster $a$. We assign to each cluster an independent random sign $S_a=\pm1$ with equal probability and define the pseudo-magnetization $M=\sum_a S_a C_a$. The rescaled variable is $x_m=|M|/\langle |M|\rangle$. This auxiliary observable probes fluctuations of signed critical cluster masses, in close analogy with the Edwards--Sokal representation of spin models. The imposed $Z_2$ symmetry makes the distribution even, so the leading $\exp(-c x^3)$ behavior concerns the positive large-$x_m$ tail, while the central part can contain additional analytic corrections.

In finite-dimensional critical systems, the full order-parameter distribution is more intricate. For example, the two- and three-dimensional Ising magnetization distribution has a double-peak structure at criticality~\cite{physreva.25.1699}, even though the transition is continuous in the thermodynamic limit. Nevertheless, its far tail is expected to obey the asymptotic form~\cite{Bruce_1995,stella25,Hilfer2003,MalakisFytas2006}
\begin{equation}
\widetilde{P}(x)
\sim
\exp(-c x^{\delta+1}),
\qquad
\delta+1=\frac{1}{1-y_h^*}.
\label{eq:mpdf}
\end{equation}
The exponent is universal, whereas the amplitude $c$ is generally nonuniversal before fixing a normalization convention. We do not assume a generic algebraic prefactor for this tail; indeed, in the complete-graph Ising limit the limiting distribution is purely quartic, $P(x)\propto \exp(-c x^4)$, without such a prefactor. Possible subleading corrections depend on the model, normalization, and boundary conditions, and require a separate analysis~\cite{Balog2025SciPost,Camia2025PowerLaw}.

The connection between probability theory and renormalization-group theory has a long history. Non-Gaussian limiting distributions for sums of strongly correlated variables were studied from the viewpoint of probabilistic limit theorems~\cite{ellis1978}. In hierarchical models, the recursion relations for probability distributions can be interpreted as real-space RG transformations, as in the work of Bleher and Sinai~\cite{bleher1973}, and in subsequent developments by Jona-Lasinio and collaborators~\cite{jona-lasinio1974,jona-lasinio1978}. More recently, functional renormalization-group methods have been used to compute critical order-parameter probability distributions and their associated rate functions~\cite{Balog2022FRG}. We summarize the fixed-point argument leading to Eq.~\eqref{eq:mpdf} in Appendix~\ref{sec:AppendixA}. A derivation based on an auxiliary cumulant-generating function~\cite{stella25} fixes the same leading stretched-exponential exponent and clarifies the origin of nonuniversal amplitudes. Its moment-normalized version is discussed in Appendix~\ref{sec:appendixB}. Throughout this work, the lattice simulations use periodic boundary conditions, and we analyze the normalized random variable $x_m=|M|/\langle |M|\rangle$.

A related spectral realization of critical order-parameter fluctuations appears in interaction-correlated random matrices constructed from two-dimensional Ising configurations~\cite{Saberi2024ICRM}. In that setting, the scaled largest eigenvalue behaves as a spectral order parameter and follows the Ising magnetization across criticality, while its critical fluctuations are described by extreme-value statistics. Thus, the same critical magnetic field can be probed spectrally, through the largest eigenvalue of an Ising-derived matrix, or directly, through the large-deviation tails of real-space order parameters studied here.

We now extend this large-deviation viewpoint to a geometric order parameter: the largest critical cluster. For percolation, the natural finite-size order parameter is the largest-cluster fraction $C_1/N$. At criticality, $\langle C_1\rangle\sim L^{y_h}$, and we define $x_C=C_1/\langle C_1\rangle$. For the Ising model, we use the Fortuin--Kasteleyn representation~\cite{fortuin1972536}, in which bonds between neighboring parallel spins are occupied with probability $p=1-e^{-2K}$ and the connected components of occupied bonds define FK clusters. Since the largest FK cluster also scales as $\langle C_1\rangle\sim L^{y_h}$ at criticality, the same normalized variable $x_C=C_1/\langle C_1\rangle$ applies to both percolation and FK-Ising criticality.

Several exact and rigorous results constrain the possible tails of $x_C$. For two-dimensional critical percolation, the right tail of the largest cluster is known, in the sense of matching upper and lower large-deviation bounds, to decay as $\exp(-c x^{96/5})$~\cite{kiss20142dper}. For the near-critical planar FK-Ising model, related estimates for the largest renormalized cluster area have also been established~\cite{camia2020}.

Let $p_C(x)$ denote the PDF of $x_C$, and define the corresponding cumulative distribution function as
\begin{equation}
F_C(x)=\Pr(x_C\le x)=\int_0^x p_C(x'),dx' .
\end{equation}
For critical percolation on the complete graph, the two tails of the largest-cluster distribution are known exactly in the scaling limit~\cite{pittel2001237}:
\begin{subequations}
\label{Eq:Fc-CGper}
\begin{align}
F_C(x)
&\sim
b_0\exp(-b_1 x^{-3/2}),
\qquad
x\to0 ,
\\
1-F_C(x)
&\sim
c_0 x^{-3/2}\exp(-c_1 x^3),
\qquad
x\to\infty .
\end{align}
\end{subequations}
This exact complete-graph result is consistent with numerical evidence for the complete graph and for seven-dimensional percolation~\cite{huang18per7d}, apart from a reported discrepancy in $b_0$ which is most likely due to a typo in Ref.~\cite{pittel2001237}. Since $y_h^*=2/3$ for complete-graph percolation, Eq.~\eqref{Eq:Fc-CGper} motivates the following scaling template for general critical clusters:
\begin{subequations}
\label{eq1}
\begin{align}
F_C(x)
&\sim
b_0\exp(-b_1 x^{-1/y_h^*}),
\qquad
x\to0 ,
\label{eq1a}
\\
1-F_C(x)
&\sim
c_0 x^{-1/y_h^*}
\exp[-c_1x^{1/(1-y_h^*)}],
\quad
x\to\infty .
\label{eq1b}
\end{align}
\end{subequations}
Here the powers of $x$ are fixed by the magnetic scaling dimension through $y_h^*=y_h/d$. For complete-graph percolation, Eq.~\eqref{eq1} reduces to the exact scaling-limit result in Eq.~\eqref{Eq:Fc-CGper}. For finite-dimensional percolation and FK-Ising clusters, we test Eq.~\eqref{eq1} as a scaling form for the normalized largest-cluster distribution. Since we use $x_C=C_1/\langle C_1\rangle$, the leading metric factor in the cluster size is removed; the remaining coefficients are dimensionless constants of the normalized scaling function and should be compared at fixed universality class, boundary conditions.

The left tail in Eq.~\eqref{eq1a} follows from a simple scaling argument. For $x_C\ll1$, no cluster reaches the typical scale $L^{y_h}$. Introduce a length $\xi\ll L$ such that the largest clusters are only of order $\xi^{y_h}$, and divide the system into $N_\xi=(L/\xi)^d$ approximately independent patches. Since $x=(\xi/L)^{y_h}$, one has $N_\xi=x^{-1/y_h^*}$. Requiring all patches to remain below their typical largest-cluster scale gives
\begin{equation}
F_C(x)
\simeq
\left[
\Pr(x_C^{(j)}\le1)
\right]^{N_\xi}
\sim
\exp(-b_1 x^{-1/y_h^*}),
\quad
x\to0 .
\end{equation}
For the right tail, $x_C\gg1$, the event $C_1\sim xL^{y_h}$ requires a coherent fluctuation of order-parameter mass on the system scale. The exponential cost is expected to be governed by the same singular exponent that controls the magnetization tail, $x^{1/(1-y_h^*)}$, while the complete-graph result supplies the algebraic prefactor $x^{-1/y_h^*}$, leading to Eq.~\eqref{eq1b}.

The contribution of this work is threefold. First, we test the magnetization-tail prediction Eq.~\eqref{eq:mpdf} for both Ising and percolation-type signed cluster-mass variables in two dimensions, three dimensions, and on the complete graph. Second, we propose and numerically test the two-tail scaling form Eq.~\eqref{eq1} for the largest-cluster CDF. Third, we identify a distinct complete-graph FK-Ising left-tail regime in which rare configurations follow percolation-like scaling, $C_1\sim N^{2/3}$, rather than the typical Ising scaling, $C_1\sim N^{3/4}$~\cite{fang21}.

The remainder of the paper is organized as follows. In Sec.~\ref{sec:simulation}, we describe the models, algorithms, and measured observables. In Sec.~\ref{sec:results}, we present the numerical results for the two-dimensional lattice, the three-dimensional lattice, and the complete graph. In Sec.~\ref{sec:conclusion}, we summarize the results and discuss open theoretical questions.

\section{Models and Sampled Quantities}
\label{sec:simulation}

In this section, we describe the models, simulation methods, and observables used to test the large-deviation tails of the order-parameter distributions.

\subsection{Model and algorithm}

We study critical bond percolation and the Fortuin--Kasteleyn Ising model on three geometries: the two-dimensional square lattice, the three-dimensional simple cubic lattice, and the complete graph. Periodic boundary conditions are used for the two lattice geometries. The FK-Ising model is simulated using the Swendsen--Wang cluster algorithm. For standard bond percolation, occupied-bond configurations are generated independently at the corresponding critical bond probability $p_c$.

For percolation and FK-Ising simulations on the complete graph with $N$ sites, the number of possible edges is $|E|=N(N-1)/2$. Directly visiting all edges is inefficient because the relevant bond probabilities scale as $O(1/N)$, so the graph is sparse at criticality. We therefore use an accelerated bond-placement method that jumps directly from one occupied bond to the next~\cite{physreve.66.066110,physreve.72.016126}. If $p$ is the bond-occupation probability and $r\in(0,1]$ is a uniform random number, the spacing to the next occupied bond is generated by
\begin{equation}
i=
1+
\left\lfloor
\frac{\ln r}{\ln(1-p)}
\right\rfloor .
\end{equation}
Here $\lfloor\cdot\rfloor$ denotes the floor function. This reduces the expected number of visited edges from $|E|$ to $p|E|$, which is essential for large complete-graph simulations.

We simulate critical percolation and FK-Ising models on complete graphs with $N=2^{16},2^{18},2^{20}$ sites, on three-dimensional lattices with $L=32,64,128$, and on two-dimensional lattices with $L=256,512,1024$. For each model and system size, at least $10^7$ samples are generated. This large statistics is needed because the right tails decay rapidly in the large-deviation regime.

\begin{figure*}[t]
    \centering
    \includegraphics[width=0.96\textwidth]{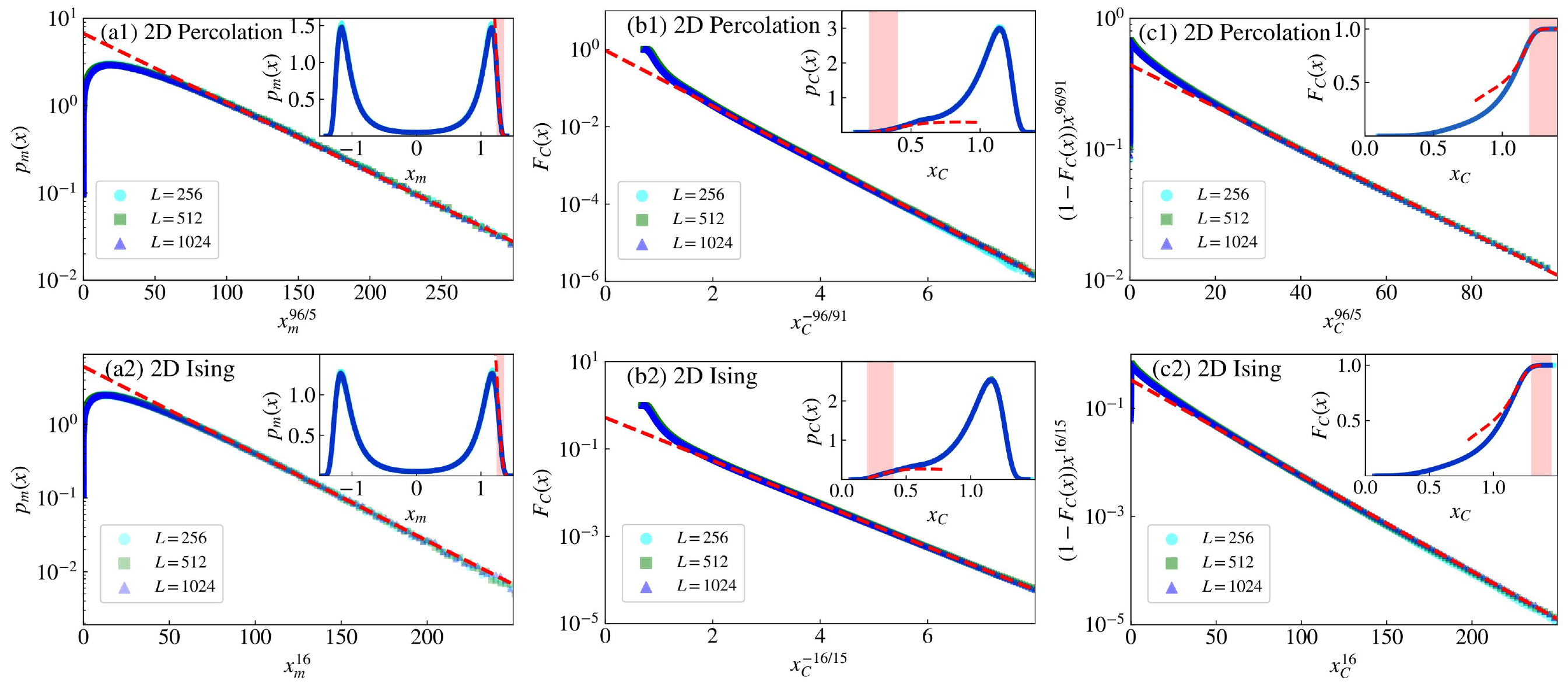}
\caption{Probability distributions of the rescaled order parameters in two dimensions. Panels (a1)--(c1) show 2D percolation, and panels (a2)--(c2) show the 2D FK-Ising model. Different colors correspond to different system sizes. Panels (a1) and (a2) show \(p_m(x)\) versus \(x_m^{1/(1-y_h^*)}\) in the large-\(x\) regime; the insets show the full PDFs of \(x_m\). Panels (b1) and (b2) show the left tail of the CDF \(F_C(x)\) versus \(x_C^{-1/y_h^*}\); the insets show the full PDFs of \(x_C\). Panels (c1) and (c2) show \((1-F_C(x))x^{1/y_h^*}\) versus \(x_C^{1/(1-y_h^*)}\) in the right tail; the insets show the full CDFs of \(x_C\). The red shaded regions indicate the fitting windows, and the dashed curves show the corresponding fits.}
\label{fig:2d}
\end{figure*}

\subsection{Sampled quantities}

For both percolation and FK-Ising models at criticality, we sample two order-parameter variables.

\begin{enumerate}[label=(\alph*)]
    \item The largest cluster size $C_1$. At criticality, its ensemble average scales as $\langle C_1\rangle\sim L^{y_h}$, equivalently as $\langle C_1\rangle\sim N^{y_h^*}$ with $y_h^*=y_h/d$. We define the rescaled largest-cluster variable
    \begin{equation}
    x_C=
    \frac{C_1}{\langle C_1\rangle},
    \end{equation}
    and analyze its probability distribution and cumulative distribution.

    \item The magnetization-like variable $M$. For the FK-Ising model, the magnetization is the usual spin sum $M=\sum_i s_i$, with $s_i=\pm1$. For percolation, where there is no microscopic spin variable, we define a signed cluster-mass analogue: each cluster $i$ is assigned an independent random sign $S_i=\pm1$ with equal probability, and
    \begin{equation}
    M=
    \sum_i S_i C_i .
    \end{equation}
    In both cases, the distribution is $Z_2$ symmetric, and we study the folded, moment-normalized variable
    \begin{equation}
    x_m=
    \frac{|M|}{\langle |M|\rangle}.
    \end{equation}
\end{enumerate}

We denote the PDFs of $x_m$ and $x_C$ by $p_m(x)$ and $p_C(x)$, respectively. The corresponding cumulative distribution functions are denoted by $F_m(x)$ and $F_C(x)$.

\section{Numerical Results}
\label{sec:results}
In this section, we present the numerical results for the probability distributions of the two rescaled order parameters, \(x_m=|M|/\langle |M|\rangle\) and \(x_C=C_1/\langle C_1\rangle\). The results are organized by geometry: the two-dimensional square lattice in Sec.~\ref{subsec:2D}, the three-dimensional simple cubic lattice in Sec.~\ref{subsec:3D}, and the complete graph in Sec.~\ref{subsec:CG}. For each geometry, we study both critical bond percolation and the critical FK-Ising model.

The tail exponents are fixed by the reduced magnetic exponent \(y_h^*=y_h/d\). As the dimension increases from two to three and then to the complete-graph limit, \(y_h^*\) decreases, and the right-tail exponent \(1/(1-y_h^*)\) decreases accordingly. This makes the large-deviation tails progressively less steep. The shape of the full distribution also changes with geometry. In low dimensions, the magnetization-like distribution \(p_m(x)\) displays the familiar double-peak structure, while in the complete-graph limit the distribution becomes single-peaked. This trend is reflected in the critical Binder ratio defined in Eq.~\eqref{eq:binder}. For the Ising universality class, \(Q=0.856(2)\) in two dimensions~\cite{kamieniarz_1993}, \(Q=0.62338(8)\) in three dimensions~\cite{deng03}, and \(Q=\Gamma^2(3/4)/[\Gamma(5/4)\Gamma(1/4)]\approx0.4569\) on the complete graph.

The analysis below focuses on the large-deviation tails predicted in Eqs.~\eqref{eq:mpdf} and~\eqref{eq1}. For \(x_m\), we test the stretched-exponential tail of the PDF. For \(x_C\), we analyze separately the left and right tails of the cumulative distribution \(F_C(x)\).

\begin{figure*}[t]
    \centering
    \includegraphics[width=0.96\textwidth]{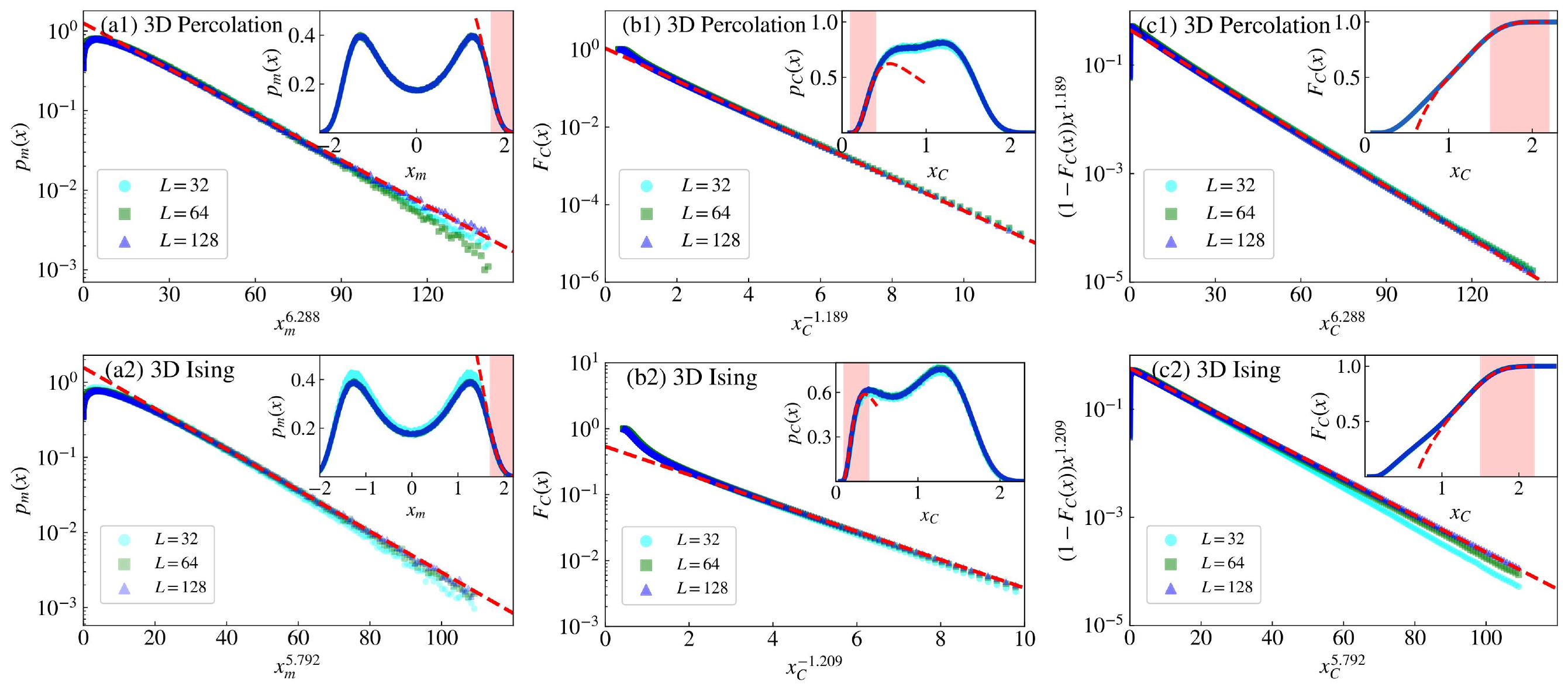}
\caption{Probability distributions of the rescaled order parameters in three dimensions. Panels (a1)--(c1) show 3D percolation, and panels (a2)--(c2) show the 3D FK-Ising model. Different colors correspond to different system sizes. Panels (a1) and (a2) show \(p_m(x)\) versus \(x_m^{1/(1-y_h^*)}\) in the large-\(x\) regime; the insets show the full PDFs of \(x_m\). Panels (b1) and (b2) show the left tail of the CDF \(F_C(x)\) versus \(x_C^{-1/y_h^*}\); the insets show the full PDFs of \(x_C\). Panels (c1) and (c2) show \((1-F_C(x))x^{1/y_h^*}\) versus \(x_C^{1/(1-y_h^*)}\) in the right tail; the insets show the full CDFs of \(x_C\). The red shaded regions indicate the fitting windows, and the dashed curves show the corresponding fits.}
\label{fig:3d}
\end{figure*}

\subsection{2D lattice}
\label{subsec:2D}

Figure~\ref{fig:2d} shows the results for the two-dimensional square lattice. The reduced magnetic exponent is \(y_h^*=15/16\) for the 2D Ising universality class and \(y_h^*=91/96\) for 2D percolation. These values imply large right-tail exponents, \(1/(1-y_h^*)=16\) for Ising and \(96/5\) for percolation, making the two-dimensional tails particularly steep.

For the magnetization-like variable \(x_m\), Figs.~\ref{fig:2d}(a1) and \ref{fig:2d}(a2) plot \(p_m(x)\) against \(x_m^{1/(1-y_h^*)}\) on a semi-logarithmic scale. The approximately linear behavior in the fitting windows supports the stretched-exponential tail predicted by Eq.~\eqref{eq:mpdf}. The insets show the full PDFs on linear axes. Both the percolation pseudo-magnetization and the FK-Ising magnetization display a double-peak structure in the full distribution.

For the largest-cluster variable \(x_C\), Figs.~\ref{fig:2d}(b1) and \ref{fig:2d}(b2) show the left tail of \(F_C(x)\) plotted against \(x_C^{-1/y_h^*}\). The observed linear behavior on the semi-logarithmic scale supports the left-tail form in Eq.~\eqref{eq1}. The right-tail test is shown in Figs.~\ref{fig:2d}(c1) and \ref{fig:2d}(c2), where \((1-F_C(x))x^{1/y_h^*}\) is plotted against \(x_C^{1/(1-y_h^*)}\). The data are consistent with the right-tail form in Eq.~\eqref{eq1}. Thus, the two-dimensional data support the proposed large-deviation scaling forms for both order parameters.

\subsection{3D lattice}
\label{subsec:3D}

Figure~\ref{fig:3d} shows the results for the three-dimensional simple cubic lattice. We use the high-precision estimates \(y_h=2.4818(4)\) for the 3D Ising universality class~\cite{3disinghou19} and \(y_h=2.522~95(15)\) for 3D percolation~\cite{junfeng13}. These give \(y_h^*\approx0.8273\) for Ising and \(y_h^*\approx0.8410\) for percolation.

For \(x_m\), Figs.~\ref{fig:3d}(a1) and \ref{fig:3d}(a2) plot \(p_m(x)\) against \(x_m^{1/(1-y_h^*)}\). The data show the expected semi-logarithmic linear trend in the large-\(x\) fitting range, supporting Eq.~\eqref{eq:mpdf}. The insets show the full PDFs. As in two dimensions, both the percolation pseudo-magnetization and the FK-Ising magnetization exhibit a double-peak structure.

For \(x_C\), Figs.~\ref{fig:3d}(b1) and \ref{fig:3d}(b2) show the left tail of \(F_C(x)\) plotted against \(x_C^{-1/y_h^*}\). The data support the left-tail form in Eq.~\eqref{eq1}. The right-tail test is shown in Figs.~\ref{fig:3d}(c1) and \ref{fig:3d}(c2), where \((1-F_C(x))x^{1/y_h^*}\) is plotted against \(x_C^{1/(1-y_h^*)}\). The observed linear trend supports the right-tail form in Eq.~\eqref{eq1}. Overall, the three-dimensional data provide further evidence that the same large-deviation scaling structure applies beyond two dimensions.

\subsection{Complete graph (CG)}
\label{subsec:CG}

\begin{figure*}
    \centering
    \includegraphics[width=0.95\textwidth]{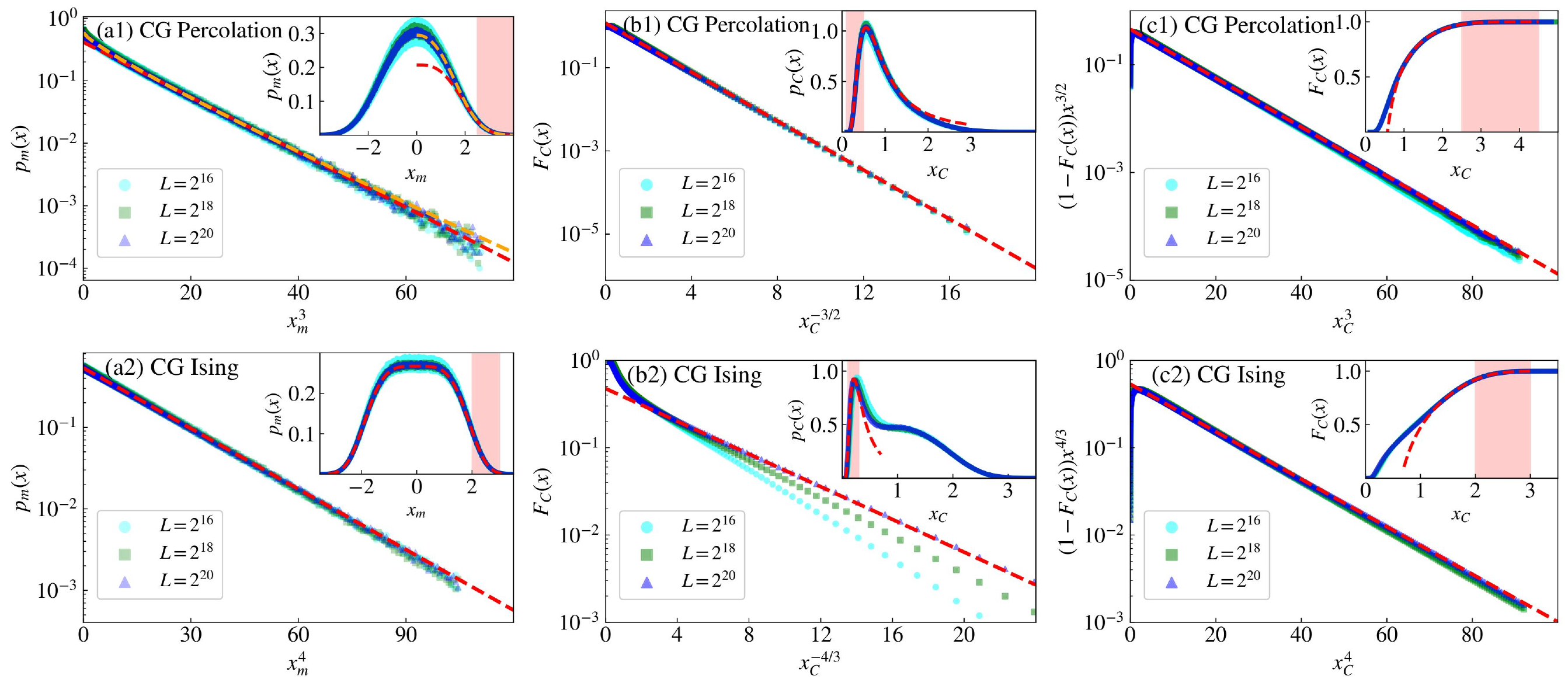}
\caption{Probability distributions of the rescaled order parameters on the complete graph. Panels (a1)--(c1) show critical percolation, and panels (a2)--(c2) show the FK-Ising model. Different colors correspond to different system sizes. Panels (a1) and (a2) show $p_m(x)$ versus $x_m^{1/(1-y_h^*)}$ in the large-$x$ regime; the insets show the full PDFs of $x_m$. Panels (b1) and (b2) show the left tail of the CDF $F_C(x)$ versus $x_C^{-1/y_h^*}$; the insets show the full PDFs of $x_C$. Panels (c1) and (c2) show $(1-F_C(x))x^{1/y_h^*}$ versus $x_C^{1/(1-y_h^*)}$ in the right tail; the insets show the full CDFs of $x_C$. The red shaded regions indicate the fitting windows, the dashed curves show the corresponding fits and the orange line in panel (a1) represents the fit of the pure stretched-exponential form with a Gaussian correction.}
\label{fig:cg}
\end{figure*}

\begin{figure}[!b]
    \centering
    \includegraphics[width=0.35\textwidth]{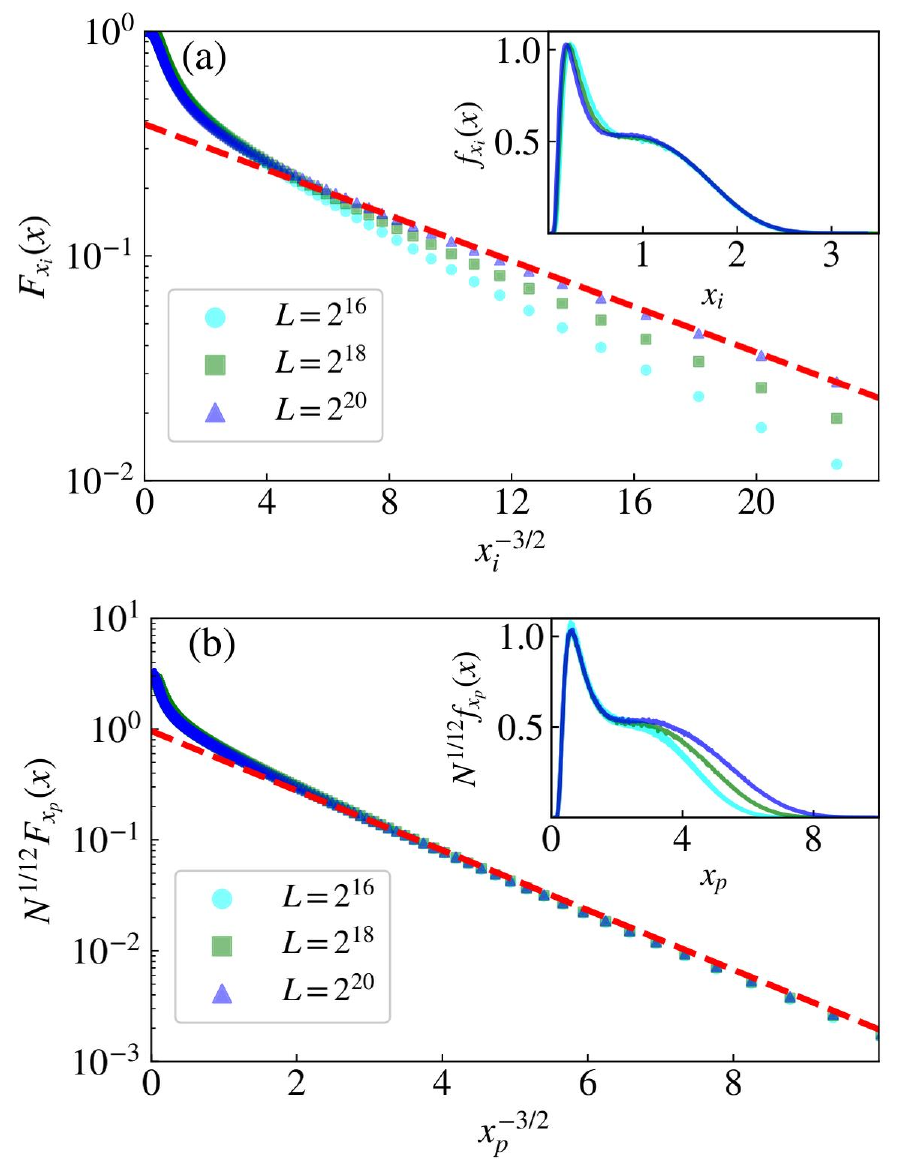}
\caption{Left-tail scaling of the largest FK cluster on the complete graph. Panel (a) shows the CDF $F_{x_i}$ of the Ising-scaled variable $x_i=C_1/N^{3/4}$. Panel (b) shows the rescaled CDF $N^{1/12}F_{x_p}$ of the percolation-scaled variable $x_p=C_1/N^{2/3}$. The main panels plot the CDFs against $x^{-3/2}$ on a semi-logarithmic scale, and the insets show the corresponding PDFs.}
\label{fig:xixp}
\end{figure}

We now turn to the complete graph, which realizes the mean-field limit. The reduced magnetic exponent is $y_h^*=2/3$ for percolation and $y_h^*=3/4$ for the Ising universality class. Figure~\ref{fig:cg} shows the corresponding distributions. Panels (a1)--(c1) show critical percolation, while panels (a2)--(c2) show the FK-Ising model.

For the magnetization-like variable $x_m$, Figs.~\ref{fig:cg}(a1) and \ref{fig:cg}(a2) plot $p_m(x)$ against $x_m^{1/(1-y_h^*)}$. For the complete-graph FK-Ising model, the data follow the mean-field quartic limiting form over the resolved range, as expected from the Landau theory discussed in the Introduction. For complete-graph percolation, the leading large-$x_m$ behavior is consistent with the mean-field percolation tail. The central part of the pseudo-magnetization distribution shows visible deviations from a pure stretched-exponential form, which is associated with the signed cluster-mass construction and its imposed $Z_2$ symmetry. Empirically, the fully resolved range is well described by adding a Gaussian correction factor to the leading tail. We treat this correction as a finite-$x_m$ feature of the auxiliary pseudo-magnetization, while the large-$x_m$ tail is the relevant test of Eq.~\eqref{eq:mpdf}.

For the largest-cluster variable $x_C$, Figs.~\ref{fig:cg}(b1) and \ref{fig:cg}(b2) show the left tail of $F_C(x)$ plotted against $x_C^{-1/y_h^*}$. Figure~\ref{fig:cg}(b1) is consistent with the complete-graph percolation result that motivated Eq.~\eqref{eq1}. Figure~\ref{fig:cg}(b2), however, shows stronger finite-size effects for the FK-Ising model. The right-tail test is shown in Figs.~\ref{fig:cg}(c1) and \ref{fig:cg}(c2), where $(1-F_C(x))x^{1/y_h^*}$ is plotted against $x_C^{1/(1-y_h^*)}$. The data support the right-tail scaling form in Eq.~\eqref{eq1} for both complete-graph percolation and complete-graph FK-Ising.

The left tail of the complete-graph FK-Ising model requires a more detailed interpretation. The typical largest FK cluster at criticality scales as $N^{3/4}$, corresponding to Ising mean-field scaling. However, rare configurations exist in which the largest cluster instead scales as $N^{2/3}$, as in critical random-graph percolation~\cite{fang21}. These rare configurations have probability of order $N^{-1/12}$. Therefore, the small-$x_C$ tail of the FK-Ising largest-cluster distribution probes the crossover from typical Ising scaling to percolation-like cluster scaling.

To make this point explicit, we introduce two rescaled variables,
\begin{equation}
x_i=\frac{C_1}{N^{3/4}},
\qquad
x_p=\frac{C_1}{N^{2/3}},
\end{equation}
corresponding to Ising and percolation scaling, respectively. Let $F_{x_i}(x)=\Pr(x_i\le x)$ and $F_{x_p}(x)=\Pr(x_p\le x)$ be their CDFs, with PDFs denoted by $f_{x_i}(x)$ and $f_{x_p}(x)$.

Figure~\ref{fig:xixp} compares these two scalings. In panel (a), $F_{x_i}$ is plotted against $x_i^{-3/2}$. In panel (b), $N^{1/12}F_{x_p}$ is plotted against $x_p^{-3/2}$. The collapse in panel (b) shows that the small-$x_C$ probability is naturally described by the percolation scale $C_1\sim N^{2/3}$, together with the probability weight $N^{-1/12}$. This confirms that the left tail of the complete-graph FK-Ising distribution is controlled by rare configurations with percolation-like cluster scaling, rather than by the typical Ising scale $N^{3/4}$.

\section{Conclusion}
\label{sec:conclusion}

We have studied large-deviation tails of critical order-parameter distributions in percolation and FK-Ising models on the square lattice, the simple cubic lattice, and the complete graph. The analysis focused on two rescaled variables: the magnetization-like variable $x_m=|M|/\langle |M|\rangle$ and the largest-cluster variable $x_C=C_1/\langle C_1\rangle$. Together, these observables probe two complementary aspects of critical fluctuations: signed order-parameter fluctuations and geometric cluster fluctuations.

For $x_m$, our simulations support the stretched-exponential large-deviation tail in Eq.~\eqref{eq:mpdf}. This behavior is observed not only for the FK-Ising magnetization, but also for the signed cluster-mass analogue introduced for percolation. On the complete graph, the FK-Ising data follow the mean-field quartic form, while the percolation pseudo-magnetization displays the expected large-$x_m$ tail together with finite-$x_m$ corrections associated with the imposed $Z_2$ symmetry of the auxiliary construction.

For the largest-cluster variable $x_C$, we proposed and tested the two-tail scaling form in Eq.~\eqref{eq1}. The left tail describes the probability that all clusters remain much smaller than the typical critical scale, while the right tail describes rare configurations with an unusually large critical cluster. The numerical data in two dimensions, three dimensions, and on the complete graph support this scaling structure when the appropriate universality-class value of $y_h^*=y_h/d$ is used. Thus, the largest-cluster distribution provides a geometric realization of the same large-deviation viewpoint that underlies the magnetization-tail scaling.

The complete-graph FK-Ising model shows a particularly informative left-tail regime. Although the typical largest FK cluster scales as $N^{3/4}$, the far left tail is governed by rare configurations in which the largest cluster follows the percolation-like scale $N^{2/3}$. The collapse of the rescaled distribution with the corresponding probability weight $N^{-1/12}$ makes this rare regime visible directly in the largest-cluster CDF.

Overall, our results show that distribution tails contain universal information that is not captured by averaged observables alone. The magnetization-like tail and the largest-cluster tails provide two related probes of rare critical fluctuations, with exponents controlled by the magnetic scaling dimension and dimensionless amplitudes fixed by the chosen normalized scaling function at fixed boundary conditions. A more complete analytical theory of the largest-cluster tails, including subleading prefactors and finite-size corrections, remains an interesting direction for future work.

\section{Acknowledgments}
This work was supported by the National Natural Science Foundation of China (under Grant No.~12275263), the Innovation Program for Quantum Science and Technology (under Grant No.~2021ZD0301900), and the Natural Science Foundation of Fujian Province of China (under Grant No.~2023J02032). The authors also thank Sheng Fang for valuable discussions and support. This work was also funded by the German Research Foundation (DFG) under Project No.~557852701 (A.A.S.) and supported by the Advanced Study Group ``Strongly Correlated Extreme Fluctuations'' at the Max Planck Institute for the Physics of Complex Systems, Dresden (2024/25)~\cite{pks_asg2024}.

\appendix
\setcounter{section}{0}
\setcounter{figure}{0}
\setcounter{equation}{0}
\setcounter{table}{0}
\renewcommand{\thefigure}{\Alph{section}\arabic{figure}}
\renewcommand{\thetable}{\Alph{section}\arabic{table}}
\renewcommand{\theequation}{\Alph{section}\arabic{equation}}

\section{Fixed-Point Analysis for the Universal Tail of \texorpdfstring{$p_m$}{p\_m}}
\label{sec:AppendixA}
\setcounter{equation}{0}

In this appendix, we first recall the fixed-point mechanism in hierarchical models and then give a simple two-block scaling argument for the leading large-deviation exponent of the critical magnetization distribution. The purpose is to justify the exponent in Eq.~\eqref{eq:mpdf}. The argument fixes the stretched-exponential power of the tail.Amplitudes and other subleading corrections depend on the normalization of the order parameter, boundary conditions, and metric factors.

The connection between real-space renormalization and probability distributions can be made explicit in Dyson's hierarchical model~\cite{bleher1973,koch1994nontrivial,meurice2007nonlinear}. In this model, the exact recursion relation for the rescaled probability distribution $f(\phi)$ is
\begin{equation}
\label{eq:hierarchical}
\begin{aligned}
\left(\mathcal{R}[f]\right)(\phi)
&=
\mathcal{N}
e^{\beta\phi^2/2}
\int_{-\infty}^{\infty}d\xi\,
f\left(\frac{\phi}{\sqrt{c}}+\xi\right)
f\left(\frac{\phi}{\sqrt{c}}-\xi\right),
\end{aligned}
\end{equation}
where $c=2^{1-2/D}$ encodes the effective dimension $D$, and $\mathcal{N}$ is a normalization constant. The Gaussian factor can be absorbed by defining $\tilde{f}(\phi)=f(\phi)e^{-A\phi^2}$ with $A=\beta c/[2(2-c)]$. This gives
\begin{equation}
\label{eq:RG_hierarchical2}
\begin{aligned}
\left(\mathcal{R}[\tilde{f}]\right)(\phi)
&=
\mathcal{N}
\int_{-\infty}^{\infty}d\xi\,
e^{2A\xi^2}
\tilde{f}\left(\frac{\phi}{\sqrt{c}}+\xi\right)
\tilde{f}\left(\frac{\phi}{\sqrt{c}}-\xi\right).
\end{aligned}
\end{equation}
At criticality, the recursion has a nontrivial fixed-point distribution $\tilde{f}^*$. For large $|\phi|$, the integral is dominated by the saddle point $\xi=0$, and the fixed-point condition reduces asymptotically to
\begin{equation}
\tilde{f}^*(\phi)
\propto
\left[
\tilde{f}^*\left(\frac{\phi}{\sqrt{c}}\right)
\right]^2 .
\end{equation}
Substituting the ansatz $\tilde{f}^*(\phi)\sim\exp(-C|\phi|^p)$ gives $p=\delta+1$, with $\delta=2\ln2/\ln c-1=(D+2)/(D-2)$. Thus
\begin{equation}
\tilde{f}^*(\phi)
\sim
\exp\left(-C|\phi|^{\delta+1}\right),
\qquad
|\phi|\to\infty .
\end{equation}

For a general critical system, an exact real-space recursion is not available, but the same tail mechanism can be formulated at the level of a two-block transformation. Consider a system with $N$ sites split into two equal subblocks with magnetizations $M_1$ and $M_2$, so that $M=M_1+M_2$. Introduce the signed scaling variables $x=M/N^{y_h^*}$ and $x_i=M_i/(N/2)^{y_h^*}$ for $i=1,2$, where $y_h^*=y_h/d$. They satisfy $x=(x_1+x_2)/2^{y_h^*}$.

Let $f(x)$ be the fixed-point distribution of the signed scaling variable. The two subblocks are coupled by an effective interaction $\mathcal{H}_{\rm eff}(M_1,M_2)$, and the induced transformation of the probability distribution can be written schematically as
\begin{align}
\label{eq:Ref}
\mathcal{R}[f](x)
&=
\mathcal{N}
\int dx_1\,
f(x_1)
f(2^{y_h^*}x-x_1)
\nonumber\\
&\quad\times
\exp\left[
-\beta
\mathcal{H}_{\rm eff}
\left(
(N/2)^{y_h^*}x_1,
(N/2)^{y_h^*}x_2
\right)
\right],
\end{align}
where $x_2=2^{y_h^*}x-x_1$. By exchange symmetry, the saddle point occurs at equal subblock magnetizations, namely $x_1=x_2=2^{y_h^*-1}x$. In the large-$|x|$ tail, Eq.~\eqref{eq:Ref} therefore reduces to
\begin{equation}
\label{eq:reduced}
\mathcal{R}[f](x)
\simeq
\mathcal{N}
f\left(2^{y_h^*-1}x\right)^2 .
\end{equation}
At criticality, scale invariance requires $\mathcal{R}[f](x)=f(x)$ at the fixed point. Substituting $f(x)\sim\exp(-a|x|^p)$ into Eq.~\eqref{eq:reduced} gives
$p=\frac{1}{1-y_h^*}$.
Thus the leading logarithmic tail is
\begin{equation}
\label{eq:AppendixA_tail}
f(x)
\sim
\exp\left[
-a|x|^{1/(1-y_h^*)}
\right],
\qquad
|x|\to\infty .
\end{equation}
This is the asymptotic form stated in Eq.~\eqref{eq:mpdf}. The derivation shows that the tail exponent is fixed by the magnetic scaling dimension $y_h^*$. The coefficient $a$, as well as other subleading corrections, depends on the chosen normalization, boundary conditions, and metric factors.

The main text uses the folded, moment-normalized variable $x_m=|M|/\langle |M|\rangle$. This change of variable rescales the argument of the distribution and folds the two symmetric tails onto $x_m>0$, but it does not change the stretched-exponential exponent. Therefore Eq.~\eqref{eq:AppendixA_tail} gives the leading large-$x_m$ behavior used in the numerical analysis.

\section{Derivation of the Universal Amplitude for the Rescaled Magnetization}
\label{sec:appendixB}
\setcounter{equation}{0}
In this appendix, we use the auxiliary cumulant-generator framework of Ref.~\cite{stella25} together with finite-size scaling to show how normalization by a magnetization moment removes the leading magnetic metric factor from the tail coefficient.
This complements Appendix~\ref{sec:AppendixA}, which explains the exponent, while the present appendix explains the role of normalization in the amplitude.

Following Ref.~\cite{stella25}, consider the critical magnetization distribution $p(x)$ of the scaling variable $x=M/N^{y_h^*}$. The auxiliary cumulant-generating function is
\begin{equation}
\tilde{G}(hN^{y_h/d})
=
\frac{1}{2}
\int_{-\infty}^{+\infty}dx\,
e^{xhN^{y_h/d}}p(x).
\end{equation}
The logarithm of the corresponding finite-size generator gives the difference between the free-energy density in an external field and the zero-field free-energy density. At criticality,
\begin{equation}
\label{eq:exten}
\begin{aligned}
g(h,T_c)-g(0,T_c)
&=
\lim_{N\to\infty}
\frac{\ln G_N(h,T_c)}{N}
\\
&=
A h^{d/y_h}.
\end{aligned}
\end{equation}
The amplitude $A$ is nonuniversal. The extensivity condition in Eq.~\eqref{eq:exten} implies the leading tail
\begin{equation}
p(x)
\sim
\exp(-c x^{\delta+1}),
\qquad
c=
\frac{\delta^\delta}{(\delta+1)^{\delta+1}}A^{-\delta}.
\end{equation}
Thus the coefficient $c$ inherits the nonuniversal amplitude $A$ of the singular free energy~\cite{physrevb.30.322}.

We now introduce a normalized magnetization variable. Since the critical magnetization distribution is symmetric, one should normalize by a nonzero even moment or by an absolute moment. We write this scale as
\begin{equation}
\label{eq:moment_scale}
R_k=
\langle |M|^k\rangle^{1/k},
\qquad
x_k=
\frac{M}{R_k}.
\end{equation}
For even $k$, the same scaling factor can equivalently be written as $R_k=\langle M^k\rangle^{1/k}$. The singular part of the finite-size free-energy density has the scaling form
\begin{equation}
\label{eq:fss}
f(t,h,L)
=
L^{-d}
f_s(c_1tL^{y_t},c_2hL^{y_h}),
\end{equation}
where $c_1$ and $c_2$ are nonuniversal metric factors. This scaling form implies, at criticality,
\begin{equation}
\label{eq:moment_scaling}
R_k
=
c_2 L^{y_h}\, r_k,
\end{equation}
where $r_k$ is a universal number fixed by the universal scaling function and by the chosen normalization convention. For even moments obtained from field derivatives, $r_k=[f_s^{(k)}(0,0)]^{1/k}$ up to the corresponding conventional normalization. For absolute moments, $r_k$ is defined directly by the universal fixed-point distribution. In both cases, the important point is that the only nonuniversal magnetic metric factor in $R_k$ is the same factor $c_2$.

The thermodynamic-limit free-energy density follows from the large-field behavior of the scaling function. For $t=0$, one has $f_s(0,z)\sim A_0z^{d/y_h}$ as $z\to\infty$, where $A_0$ is universal. Substituting $z=c_2hL^{y_h}$ into Eq.~\eqref{eq:fss} gives
\begin{equation}
f(0,h,\infty)
\sim
A_0 c_2^{d/y_h}h^{d/y_h},
\end{equation}
so that $A=A_0c_2^{d/y_h}$. The same metric factor $c_2$ therefore appears both in the free-energy amplitude $A$ and in the magnetization scale $R_k$.

Let $p_k(x_k)$ denote the PDF of the normalized variable $x_k=M/R_k$. Since $p_k(x_k)\,dx_k=p(M)\,dM$, the large-$|x_k|$ tail has the form
\begin{equation}
p_k(x_k)
\sim
\exp(-c_k x_k^{\delta+1}).
\end{equation}
Using Eqs.~\eqref{eq:exten}--\eqref{eq:moment_scaling}, the coefficient is
\begin{equation}
\label{eq:ck_universal}
c_k
=
\frac{\delta^\delta}{(\delta+1)^{\delta+1}}
\left(
\frac{r_k}{A_0^{y_h^*}}
\right)^{\delta+1}.
\end{equation}
All dependence on the nonuniversal metric factor $c_2$ cancels. Thus, for a fixed normalization convention, the leading tail coefficient is universal.

In the simulations and main text we use the folded first absolute moment,
$x_m=|M|/\langle |M|\rangle$.
This corresponds to the choice $k=1$ in Eq.~\eqref{eq:moment_scale}, followed by folding the symmetric distribution under $M\to |M|$. The folding changes the normalization of the PDF but not the large-deviation exponent. The leading tail therefore has the form used in Eq.~\eqref{eq:mpdf}, with an amplitude fixed by the chosen absolute-moment normalization.
\bibliography{references}
\end{document}